\documentclass[useAMS,usenatbib,usegraphicx]{mn2e}
\usepackage{epsf}
\title[Quantifying the Cosmic Web]
{Quantifying the Cosmic Web I: The large-scale halo
ellipticity-ellipticity and ellipticity-direction correlations}
\author[J. Lee, V. Springel, U.L. Pen and G. Lemson]
{Jounghun Lee$^{1}$\thanks{E-mail:jounghun@astro.snu.ac.kr}, 
Volker Springel$^{2}$, Ue-Li Pen$^{3}$, and Gerard Lemson$^{4,5}$\\
$^{1}$Department of Physics and Astronomy, FPRD, Seoul National
University, Seoul 151-747, Korea \\
$^{2}$Max-Planck-Institute for Astrophysics,
Karl-Schwarzschild-Str. 1, D-85741 Garching, Germany\\
$^{3}$Canadian Institute for Theoretical Astrophysics, Toronto, ON
M5S, Canada\\
$^{4}$Astronomisches Rechen-Institut, Zentrum fur Astronomie der Universitat
Heidelberg, Moenchhofstr. 12-14, 69120 Heidelberg, Germany\\
$^{5}$Max-Planck Institut fur extraterrestrische Physik, Giessenbach 
Str., 85748 Garching, Germany}
\begin{document}
\date{Accepted 2007 ???. Received 2007 ???; in original form 2007 September 10}
\pagerange{\pageref{firstpage}--\pageref{lastpage}} \pubyear{2007}
\maketitle
\label{firstpage}

\begin{abstract}
The formation of dark matter halos tends to occur anisotropically along
the filaments of the Cosmic Web, which induces both
ellipticity-ellipticity (EE) correlations between the shapes of halos,
as well as ellipticity-direction (ED) cross-correlations between halo
shapes and the directions to neighboring halos.  We analyze the
halo catalogue and the semi-analytic galaxy catalogue of the recent
Millennium Run Simulation to measure the EE and ED correlations numerically
at four different redshifts ($z=0$, $0.5$, $1$ and $2$). For the EE
correlations, we find that (i) the major-axis correlation is strongest
while the intermediate-axis correlation is weakest; (ii) the signal is
significant at distances out to $10\,h^{-1}$Mpc; (iii) the signal
decreases as $z$ decreases; (iv) and its behavior depends strongly on
the halo mass scale, with larger masses showing stronger correlations
at large distances. For the ED correlations, we find that 
(i) the correlations are much stronger than the EE correlations, 
and are significant even out to distances of $50\,h^{-1}$Mpc; 
(ii) the signal also decreases as $z$ decreases; (iii) and
it increases with halo mass at all distances. We also provide
empirical fitting functions for the EE and ED correlations. 
The EE correlations are found to scale linearly with the linear density 
correlation function, $\xi(r)$. While the ED cross-correlation is
found to scale as $\xi^{1/2}(r)$ at large distances beyond $10\,h^{-1}$Mpc.  
The best-fit values of the fitting parameters for the EE and the ED
correlations are all determined through $\chi^{2}$-statistics. Our results
may be useful for quantifying the filamentary distribution of dark
matter halos over a wide range of scales.
\end{abstract}


\section{Introduction}

One of the most striking features of the Universe is that the observed
distribution of galaxies on large scales shows a web-like
filamentary pattern, which is often called the "Cosmic Web".  Recent
large N-body simulations of the cold dark matter cosmology demonstrated
vividly the geometric richness of the filamentary web that spatially
connects the dark matter halos, and which directly relates to the
structure seen in the galaxy distribution.  One of the most fundamental
tasks in cosmology is thus to establish a physical model for the
filamentary cosmic web and to quantitatively explain its global properties.

The existence of the filamentary web was originally predicted by the
top-down scenario of the hot dark matter (HDM) model \citep{zel70}.  If
cosmic structures form through top-down fragmentation, then one- and
two-dimensional collapse of matter would naturally lead to the formation
of sheet-like and filamentary structures on large scales. Therefore, it
was regarded first as a mystery why and how the filamentary web came
into being also in a cold dark matter (CDM) dominated universe.

A breakthrough was made by \citet{bon-etal96} who developed a cosmic
web theory that can explain the presence of a filamentary web
in the CDM cosmogony. This theory explains that the filamentary web
can occur naturally in the CDM dominated universe due to the coherent 
nature of the primordial tidal field. The filamentary web is in fact a
manifestation of the primordial tidal field sharpened by nonlinear 
effects. The cosmic web theory has provided a standard framework
within which the formation of cosmic large-scale structure can be 
qualitatively understood.  Yet, it is still quite difficult to 
describe the cosmic web quantitatively both in theoretical and in 
observational terms. Theoretically, the inherent anisotropic nature
and geometrical complexity of the cosmic web makes it complicated to
fully characterize its statistical properties. Observationally, it is
hard to trace the filamentary structures from observational data,
since there is no well-established way to identify them.

In spite of these difficulties, various methodologies and algorithms
have already been suggested to quantify the filamentary structures:
Higher-order N-point statistics has been used to describe the
anisotropic matter distribution in a cosmic web
\citep{cro-etal04,kul-etal07}; the percolation statistics was used to
characterize the filamentary shapes of the large-scale structures
\citep{sah-etal97,sha-yes98}; the skeleton formalism has been developed
to extract the filamentary structures from a three dimensional density
field \citep[e.g.,][and references therein]{sou-etal07}; the
Minimal-Spanning-Three algorithm has been introduced to find the basic
structural elements of the cosmic web \citep{col07}.

Although the above methods are quite useful for determining the
overall filamentary structure of the cosmic web, these approaches are
largely phenomenological without accounting for the physical mechanism
for the formation of the cosmic web. According to the theory proposed by
\citet{bon-etal96}, the filamentary web originated from the large-scale
coherence of the primordial tidal field and its sharpening by nonlinear
effects during structure growth.  Part of this nonlinear sharpening
effect arises from the gravitationally driven merging of halos and the
infall of matter, which preferentially occurs along the most prominent
filaments. This increases the anisotropy in the halo clustering and thus
sharpens the filamentary web.

To describe the cosmic web quantitatively in terms of its
underlying physical principles it will be necessary to account for the
effects of the tidal field and the anisotropic merging along filaments.
The tidal field causes intrinsic alignments of the principal axes of the
dark halos in the cosmic web
\citep{cro-met00,hea-etal00,cat-etal01,jin02,hui-zha02,lee-pen07},
while anisotropic merging induces elongation of the major axes of the
halos along prominent filaments \citep{wes89,wes-etal91}.  As a result,
there exist spatial correlations between the halo ellipticities (EE
correlations), and cross-correlations between the halo ellipticities and
the large-scale density field (ED cross-correlations). Hence, one can
view the observed filamentary web as a large-scale manifestation of the
EE and ED correlations, which are in turn induced by the effects of the
tidal field and the anisotropic merging.

The goal of this paper is to quantify the filamentarity and the typical
scales of the cosmic web in terms of the EE and ED correlations. This is
also highly important for assessing to what degree these correlations
can systematically bias weak gravitational lensing mass reconstructions
and cosmological parameter estimates based on cosmic shear measurements.
In fact, the ED cross-correlations have become a hot issue in the weak
lensing community, since it has been realized that they could mimic weak
lensing signals at a significant level
\citep{hir-sel04,man-etal06,hir-etal07}.

Several authors have already studied numerically the EE and ED
correlations. \citet{hop-etal05} examined the evolution of cluster
alignments by using a large-scale N-body simulation. 
\citet{alt-etal06} have shown from high-resolution N-body data that
the alignments of clusters are strongly related to the existence of
the  connecting filaments.  \citet{hey-etal06} explored the possible
correlations between the weak lensing shear and the orientation of
foreground galaxies by analyzing N-body simulation. Here in our work, 
we measure both the EE and ED correlations at large distances by
using the Millennium dataset and describe quantitatively their
scalings with distance, mass, as well as redshift.  

The organization of this paper is as follows. In Section~2, we describe the
N-body dataset we use and explain how we measure halo ellipticities from
the N-body simulation and its associated galaxy catalogue. In Sections~3 
and 4, we report numerical detections of the EE correlations and the ED
cross-correlations, and examine how the signals depend on distance scale, 
redshift and halo mass. We also provide useful fitting formula for them.  
Finally, in Section~5, we summarize the results and discuss the implications 
of our work.

\section{Simulation data and methodology}

Our analysis is based on the halo catalog and the semi-analytic galaxy
catalog from the recent high-resolution 
{\em Millennium Simulation}\footnote{The Millennium Simulation data are  
available at http://www.mpa-garching.mpg.de/millennium}, 
which followed $10^{10}$ dark matter particles in a $\Lambda$CDM concordance
cosmology \citep{spr-etal05}. The size of the periodic simulation box
is $500\, h^{-1}$Mpc and each dark matter particle in the simulation has
a mass of $8.6\times 10^{8}h^{-1}M_{\odot}$. The basic cosmological
parameters of the simulation were chosen as $\Omega_{m}=0.25$ (the mass
density); $\Omega_{\Lambda}=0.75$ (the vacuum energy density); $h=0.73$
(the dimensionless Hubble constant); $\sigma_{8}=0.9$ (the linear power
spectrum amplitude); and $n_{s}=1$ (the slope of the primordial power
spectrum).

As part of the analysis of the Millennium run, halos of dark matter
particles were first identified with the standard friends-of-friends
(FOF) algorithm, and then decomposed into gravitationally bound subhalos
using the {\small SUBFIND} algorithm \citep{spr-etal01}.  Based on
detailed merger trees constructed for the subhalos, the halos were then
populated with luminous galaxy models using semi-analytic simulations
of the galaxy formation process \citep{cro-etal06}.

We here use the spatial distribution of subhalos and galaxies to
characterize the shape of FOF halos, in analogy to the procedure applied
to observational galaxy surveys \citep[e.g.,][]{mei-etal07}.  
For each FOF halo, we locate the satellite galaxies belonging to it. 
Then, we measure their tensor $(I_{ij})$ of second order mass moments as
\begin{equation}
I_{ij} = \sum_{\alpha}m_{\alpha}\,x_{\alpha,i}\,x_{\alpha,j},
\label{eqn:iner}
\end{equation}
where $m_{\alpha}$ is the luminosity (or, equivalently the stellar mass) 
of the $\alpha$-th galaxy and $\vec{x}_{\alpha}$ is the position of the 
$\alpha$-th galaxy measured from the center of the mass of the satellite 
galaxies.  We restrict our analysis to FOF halos massive enough to contain 
more than five substructures. By diagonalizing $I_{ij}$, we determine
the three principal axes (major, intermediate, and minor axes) of
$I_{ij}$. This allows us to measure the correlations between the three 
axes of the FOF halos as a function of separation. The results of our 
measurements are presented in detail in Section~3.

Before turning to our results, it is worth to discuss how our
methodology relates to other, previously applied methods to characterize
the shape of halos.  Note that we here do not use all dark matter
particles of a FOF halo to measure $I_{ij}$. Instead, we only use the
satellite galaxies (or substructures) as tracers of the shape.  In
general, measuring the shape of a halo is a somewhat ambiguous issue,
where a number of different strategies have been applied in the
literature, but no generally accepted standard procedure exists
\citep[see e.g. the discussion in][]{springel04,allgood06}.  
Part of the ambiguity in measuring halo shape stems from the fact that 
one cannot delineate the outer boundary of a halo in a clear-cut way.  
If all particles belonging to a FOF halo are used to measure $I_{ij}$, 
then the ellipticity of a halo may be overestimated because of the large
weight of the most distant points on the major axis, while in contrast,
if only those particles within a certain spherical radius are used to
measure $I_{ij}$, then the ellipticity of the halo is likely
underestimated.

We use satellite galaxies (or substructures) to measure $I_{ij}$,
because this approach mimics observationally accessible procedures. We
expect that this definition should give halo shapes similar to those
measured with all dark matter particles, as substructure and galaxy
density are tracers of the dark matter distribution 
\citep[e.g.,][]{agu-bra06}. We will explicitly test this below.  Note
that we focus on quantifying the filamentary web induced by the
anisotropic {\em orientations} of the halo ellipticities; we are not
really interested in the {\em magnitude} of the shape distortion
itself.  This means that we are less sensitive to the details of
measuring halo shape compared with attempts to quantify the axis ratio 
of the halo shape.

\begin{figure}
\begin{center}
\label{fig:par}
\includegraphics[width=0.95\hsize]{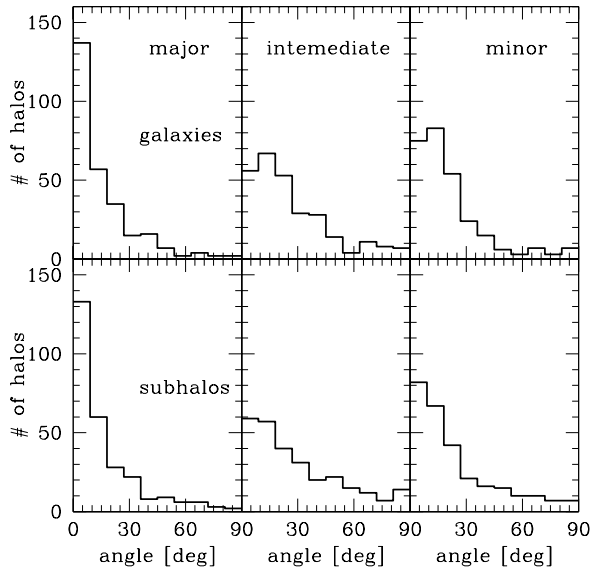}
\end{center}
\caption{Distributions of the angles between the major, intermediate,
and minor axes of the halos (left, middle, and right, respectively) from
the milli-Millennium simulation, as determined by three different
methods A, B, and C, which are based on the satellite galaxies,
subhalos, and all particles belonging to the halos, respectively. The
top-panels show the results from a comparison between the methods A and
C, while the bottom panels give a comparison between the methods B and
C.}
\label{fig:dis}
\end{figure}

To examine to what extent different measuring methods yield different
halo ellipticities, we carry out a simple test: Using a total of $227$
FOF halos from the `milli'-Millennium simulation, which is smaller test
run of the main Millennium simulation with a box size of
$62.5\,h^{-1}$Mpc \citep{spr-etal05}, we measured the halo shapes with
three different methods: (A) using galaxies weighted by luminosity; (B)
using dark matter substructures weighted by mass; and (C) using all dark
matter particles.  Then, we calculate the distribution of the angles
$\theta$ between the principal axes of the halo shapes determined by
these three methods. Figure~\ref{fig:dis} compares methods A and C
(top), and methods B and C (bottom), in both cases plotting the
histogram of the angles between the halos' major, intermediate, and
minor axes (left, middle, and right panels, respectively). As can be
seen, there is a strong peak at $\theta = 0$, demonstrating that the
halo principal axes obtained by the three different methods A, B and C
are strongly correlated with one another. This correlation is
particularly robust for the major axis, which defines the primary
orientation of the predominantly prolate halos.

\section{The Halo Ellipticity-Ellipticity Correlation}

\subsection{Definition}

We define the EE correlation function, $\eta(r)$, as
\begin{equation}
\label{eqn:1st}
\eta(r)\equiv\langle\vert\hat{\bf e}_({\bf x})\cdot
\hat{\bf e}_({\bf x}+{\bf r})\vert^{2}\rangle -\frac{1}{3}
\end{equation}
where $\hat{\bf e}\equiv (\hat{e}_{i})$ represents the normalized
eigenvector of a halo with unit magnitude. In eq.(\ref{eqn:1st})
the constant $1/3$ is subtracted since the first average term will
yield $1/3$ in case that there is no EE correlation.  Now that the
three eigenvectors of each halo in the Millennium catalogs are all 
determined by the method described in Section~2, one can measure the
EE correlations of the major ($\eta_{\rm I}$), intermediate
($\eta_{\rm II}$) and minor ($\eta_{\rm III}$) principal axes separately 
as a function of the comoving distance $r$. 

Basically, for each halo pair at a given redshift, we measure the separation 
distance $r$ between the halo centers and calculate the squares of the dot 
products of the normalized eigenvectors of two halos. Then, we bin the
radial distance $r$ and calculate the mean values of 
$\vert\hat{\bf e}_({\bf x})\cdot\hat{\bf e}_({\bf x}+{\bf
r})\vert^{2}$ averaged over those halo pairs whose separation
distances belong to a given bin, subtracting $1/3$ from it. 
We perform this procedure at $z=2,1,0.5$ and $0$. 

\subsection{Evolution with redshift}

\begin{figure}
\begin{center}
\includegraphics[width=0.95\hsize]{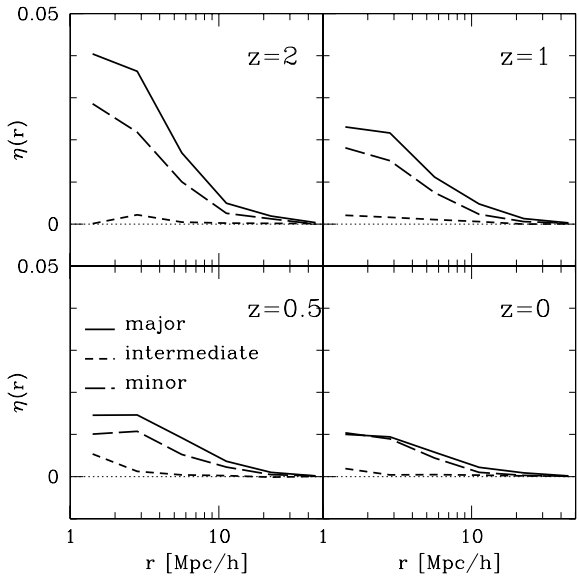}
\end{center}
\caption{EE correlations of the halo major, intermediate, and minor axes
(solid, dashed, and long dashed lines) at four different redshifts: 
$z=0$, $0.5$, $1$, and $2$ (top-left, top-right, bottom-left, and 
bottom-right panel, respectively).
\label{fig:three}}
\end{figure}
Figure~\ref{fig:three} plots the numerical results on the EE correlations 
measured at $z=0$, $0.5$, $1$ and $2$ in the top-left, top-right, bottom-left 
and bottom-right panels, respectively. In each panel, the solid, dashed and 
long dashed lines represent $\eta_{\rm I}(r)$, $\eta_{\rm II}(r)$ and 
$\eta_{\rm III}(r)$, respectively. The dotted line corresponds to the case 
of no correlation. As can be seen, the major-axis correlations are strongest 
and the intermediate-axis correlations are almost zero at all redshifts.

\begin{figure}
\begin{center}
\includegraphics[width=0.95\hsize]{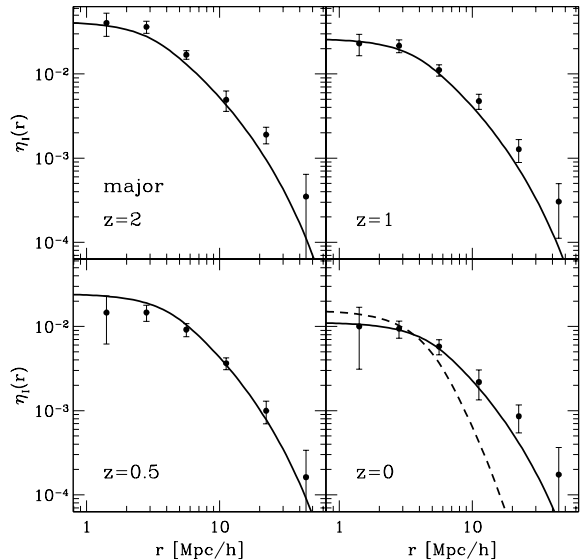}
\end{center}
\caption{The EE correlations of the halo major axes at $z=0$, $0.5$,
$1$ and $2$. The errors, $\sigma_{\eta}$, represent the standard deviation 
between $8$ realizations that are obtained from the subdivision of the 
simulation box. The errors do not include the correlations between radial 
bins. In each panel the solid line represents the fitting model 
(eq.\ref{eqn:cor}) proportional to the linear density correlation function, 
$\xi(r)$. In the bottom-right panel, the dashed line represents another 
fitting model proportional to $\xi^{2}(r)$.
\label{fig:corz}}
\end{figure}
\begin{figure}
\begin{center}
\includegraphics[width=0.95\hsize]{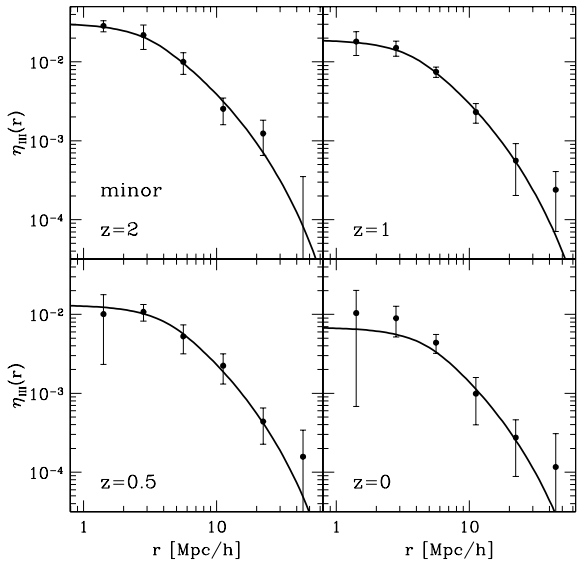}
\end{center}
\caption{Same as Fig. \ref{fig:corz} but for the case of the halo minor
axes. \label{fig:corz3}}
\end{figure}

To see the behaviors of the EE correlations at large distances, we plot 
$\eta_{\rm I}(r)$ and $\eta_{\rm III}(r)$ as solid dots with errors 
$\sigma_{\eta}$ in the logarithmic scale in Figs~\ref{fig:corz} and 
\ref{fig:corz3}, respectively. In each panel the solid line represents the 
fitting model (see Section~3.4). For the estimation of $\sigma_{\eta}$, 
we divide the simulation volume into eight subvolumes each of which has a 
linear size of $250h^{-1}$Mpc and measure the EE correlations in each 
subvolume separately. The errors, $\sigma_{\eta}$, are calculated as the 
standard deviation between realizations. This estimation of errors accounts 
for both the cosmic variance and the Poisson noise. It is also found that 
there exist non-negligible correlations between radial bins at distance 
larger than $5\,h^{-1}$Mpc (see Section~3.4).

As can be seen, the EE correlations are strongest at $z=2$, and exist
out to distances of $10h^{-1}$Mpc. As $z$ decreases, the correlations 
tend to decrease monotonically at all distance scales, indicating that 
the directions of the halo major and the minor axes tend to become randomized 
as $z$ decreases. This result is consistent with the previous results
obtained by \citet{hop-etal05}. The numerical results on the EE correlations 
measured at $z=0$ are listed in Table \ref{tab:EE_z0}.
\begin{table}
\centering
\caption{Numerical results on the EE correlations of the halo major 
($\eta_{\rm I}$) and the minor axes ($\eta_{\rm III}$) in logarithmic 
scale measured at $z=0$.
\label{tab:EE_z0}}
\begin{tabular}{@{}ccc@{}}
\hline
$\log [r/(h^{-1}{\rm Mpc})]$ & $\eta_{\rm I}(r)\times 10^{2}$ & 
$\eta_{\rm III}(r)\times 10^{2}$\\
\hline
$0.15$ & $1.00\pm 0.69$  & $1.04\pm 0.97$\\
$0.45$ & $0.94\pm 0.22$  & $0.89\pm 0.38$\\
$0.75$ & $0.58\pm 0.12$  & $0.44\pm 0.12$\\
$1.05$ & $0.22\pm 0.08$  & $0.10\pm 0.06$\\
$1.35$ & $0.09\pm 0.03$  & $0.03\pm 0.02$\\
$1.65$ & $0.02\pm 0.02$  & $0.01\pm 0.02$\\
\hline
\end{tabular}
\end{table}

\subsection{Variation with mass}

\begin{figure}
\begin{center}
\includegraphics[width=0.95\hsize]{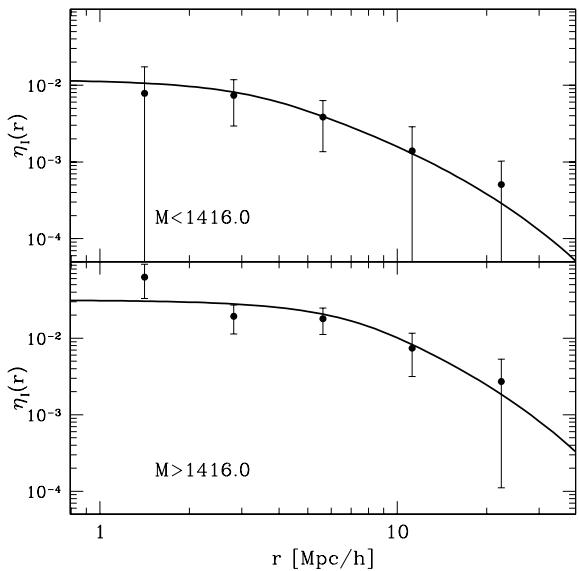}
\end{center}
\caption{The EE correlations of the halo major axes measured from the 
low-mass ($M<M_{c}$) and the high-mass ($M>M_{c}$) bins at $z=0$ in the 
top and bottom panel, respectively. The mass threshold $M_{c}=1416$ is in 
unit of $10^{10}h^{-1}M_{\odot}$.\label{fig:corm}}
\end{figure}
To study how the EE correlations scale with halo mass, we measure 
$\eta_{\rm I}(r)$ from two different mass bins with the mass threshold 
$M_{c}=1.42\times 10^{13}h^{-1}M_{\odot}$ at $z=0$. Table \ref{tab:EE_m} 
lists the numerical results and Figure~\ref{fig:corm} plots $\eta_{\rm I}(r)$ 
at $z=0$ measured from the low-mass ($M<M_{c}$) and the high-mass bin 
($M>M_{c}$) as solid dots in the top and the bottom panel, respectively. 
As can be seen, the EE correlations of the high-mass halos are stronger 
at all distances than that of the low-mass halos, which implies that 
the EE correlations increase as the halo mass increases. This finding 
is also consistent with the previous results obtained for the cluster 
alignments by \citet{hop-etal05}. 

\begin{figure}
\begin{center}
\includegraphics[width=0.95\hsize]{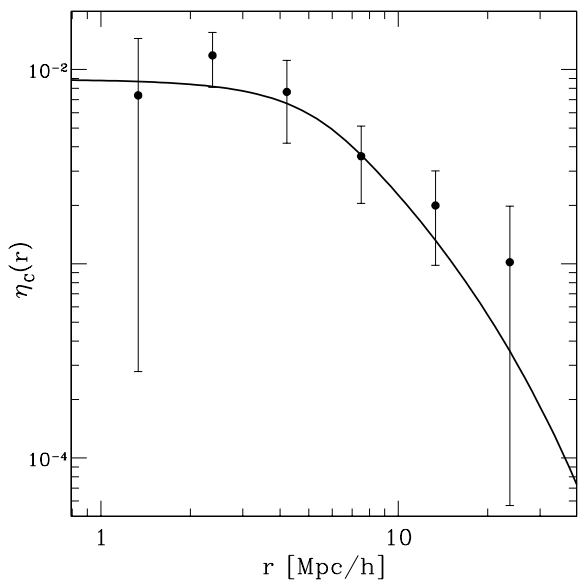}
\end{center}
\caption{The EE cross-correlations of the major axes between the high and 
the low mass halos at $z=0$.\label{fig:corc} }
\end{figure}

\begin{table}
\centering
\caption{The EE correlations of the major axes of the low-mass and 
the high-mass halos.
\label{tab:EE_m}}
\begin{tabular}{@{}ccc@{}}
\hline
$\log [r/(h^{-1}{\rm Mpc})]$ & $\eta_{\rm I}(r)\times 10^{2}$ & 
$\eta_{\rm I}(r)\times 10^{2}$ \\
&low-mass&high-mass \\
\hline
$0.15$ & $0.78\pm 0.95$  & $6.28\pm 2.98$ \\
$0.45$ & $0.74\pm 0.44$  & $1.93\pm 0.79$ \\
$0.75$ & $0.38\pm 0.25$  & $1.80\pm 0.68$ \\
$1.05$ & $0.14\pm 0.15$  & $0.74\pm 0.42$ \\
$1.35$ & $0.05\pm 0.05$  & $0.27\pm 0.26$ \\
$1.65$ & $0.01\pm 0.03$  & $0.09\pm 0.07$ \\
\hline
\end{tabular}
\end{table}
We have also measured the EE cross-correlations, $\eta_{C}(r)$, between
the low-mass and the high-mass halos. Basically, we select halo pairs each 
of which consists of one halo from the low-mass bin and one halo from 
the high-mass bin, and then measure the EE correlations of these halo pairs. 
Figure~\ref{fig:corc} plots $\eta_{C}(r)$ at $z=0$ as solid dots.
As can be seen, there exist significant cross-correlations between the
two mass bins. Note that the EE cross-correlation is in fact stronger
than the EE auto-correlation of the low-mass halos ($M<M_{c}$) but
weaker than the EE auto-correlation of the high-mass halos ($M>M_{c}$). 
This result suggests that the anisotropic merging and infall along filaments 
tend to increase not only the EE correlations on the same mass scale but 
also the EE cross-correlations between different mass scales.

\subsection{Fitting formula}

According to the first order linear model, the EE correlations are induced by 
the spatial correlations of the primordial tidal field and can be approximated 
in the linear regime by a quadratic scaling of the linear correlation function 
as $\eta(r)\propto \xi^{2}(r)$ \citep{cat-etal01,lee-pen01,hir-sel04}. 
Since $\xi^{2}(r)$ decreases rapidly with separation distance $r$, the linear 
model predicts that the EE correlations exist only between close pairs with 
$r$ less than a few Mpc. In other words, the large scale EE correlations 
cannot be described by the linear model.

\citet{hui-zha02} pointed out that the growth of the non-Gaussianity in the 
density field should cause the EE correlations to scale linearly with $\xi(r)$.
It indicates that the non-Gaussianity tends to increase the large-scale EE 
correlations. Although \citet{hir-sel04} discussed that the assumption of 
$\eta(r)\propto\xi(r)$ is valid only in the linear regime, we use here the 
following fitting formula for $\eta(r)$:
\begin{equation}
\label{eqn:cor}
\eta(r) \approx a\tilde{\xi}_{A}(r).  
\end{equation}
Here $a$ is a fitting parameter, representing the amplitude of the EE 
correlation, whose value is to be determined empirically. Since $\eta(r)$ is 
always positive and does not exceed $2/3$, the correlation parameter $a$ is 
expected to be in the range of $[0,\,2/3]$. 

In eq.~(\ref{eqn:cor}), $\tilde{\xi}(r)_{A}$ is the {\it rescaled} 
two-point correlation function of the linear density field, defined as
\begin{equation}
\label{eqn:xi}
\tilde{\xi}_{A}(r) \equiv 
\frac{\int P(k)[(\sin kr)/kr]W^{2}(k;M)\,{\rm d}^{3}k}
{\int P(k)W^{2}(k;M)\,{\rm d}^{3}k},
\end{equation}
which satisfies $\tilde{\xi}_{A}(0)=1$. Here, $P(k)$ is the linear power 
spectrum, and $W(k;M)$ is the top-hat spherical filter corresponding to the 
mass scale $M$. We have employed the approximate formula given by 
\citet{bar-etal86} for the $\Lambda$CDM power spectrum, using the same values 
of the cosmological parameters that were used for the Millennium Run 
simulation. For the shape factor $\Gamma$ of the power spectrum 
parameterization, we adopted $\Gamma =\Omega_{m}h$. For the smoothing mass 
scale $M$ in eq.~(\ref{eqn:xi}), we use the mean mass averaged over the 
selected FOF halos.

We fit the numerical results obtained in Section~3.2 to eq.~(\ref{eqn:cor}) 
by adjusting the parameter, $a$ with the help of the maximum likelihood method 
\citep{bar91}. Basically, it amounts to finding the minimum of the $\chi^{2}$ 
function defined as
\begin{equation}
\label{eqn:chi2}
\chi^{2} = [\eta_{i}-\eta(r_{i};a)]C^{-1}_{ij}
[\eta_{j} - \eta(r_{j};a)] 
\end{equation}
where  $\eta_{i}$ is the numerical data point at the $i$-th radial bin, $r_i$, 
$\eta(r_{i};a)$ represents the fitting model (eq.~\ref{eqn:cor}) calculated 
at $r_i$, and  $(C^{-1}_{ij})$ is the inverse of the covariance matrix, 
$(C_{ij})$, whose component is calculated as the ensemble average over the 
8 realizations 
\citep{bar91}:
\begin{equation}
\label{eqn:cov_eta}
C_{ij} = \langle(\eta_{i}-\eta_{0i})(\eta_{j}-\eta_{0j})\rangle, 
\end{equation} 
where $\eta_{0i}$ represents the mean $\eta_{i}$ obtained from the whole 
simulation box. It is worth mentioning here that the $\chi^{2}$ function 
is expressed in terms of the inverse covariance matrix, $(C_{ij})$, given 
that there exist non-negligible correlations between radial bins 
at distance scales larger than $5\,h^{-1}$Mpc.
In our case the number of realizations is larger than that of radial bins, 
the $\chi^{2}$ function defined in terms of the inverse covariance matrix 
should be useful to find the best-fit value of $a$ \citep{har-etal07}.

The uncertainty in the measurement of $a$ is calculated as the curvature of 
the $\chi^{2}$ function at the minimum \citep{bev-rob96}:
\begin{equation}
\sigma^{2}_{a} = \left(\frac{\partial^{2}\chi^{2}}{\partial a^{2}}\right)^{-1},
\end{equation}
The fitting results are summarized in Table \ref{tab:corz_fit} which lists the 
mean mass $\bar{M}$ in unit of $10^{10}h^{-1}M_{\odot}$, the number of halos 
$N_{h}$, and the best-fit values of $a$ at four different redshifts.  
\begin{table}
\centering
\caption{Redshift ($z$), halo mean mass ($\bar{M}$) in unit of
$10^{10}h^{-1}M_{\odot}$, number of halos ($N_{h}$), and the best-fit
values of $a$ for the EE correlations of the halo major axes.
\label{tab:corz_fit}}
\begin{tabular}{@{}ccccc@{}}
\hline
z & $\bar{M}$ & $N_{h}$ & $a\times 10^{2}$ & \\
& [$10^{10}h^{-1}M_{\odot}$] &    &  &  \\
\hline
0   & $1638.15$ & $121773$ & $1.12\pm 0.06$ &  \\
0.5 & $1118.82$ & $131505$ & $2.45\pm 0.11$ & \\
1   & $810.92$  & $125363$ & $2.64\pm 0.23$ &  \\
2   & $448.33$  & $73514$  & $3.01\pm 0.39$ &  \\ 
\hline
\end{tabular}
\end{table}
Note that the value of $a$ deviates from zero at all redshifts and decreases 
monotonically as $z$ decreases. These results quantify well how strong the EE 
correlations are and how they evolve with redshifts. The fitting models with 
these best-fit-values of $a$ are plotted as solid line in Fig.~\ref{fig:corz}.
As can be seen, the fitting models are in good agreement with the numerical 
results (solid dots) at all redshifts. For comparison, we also try to fit 
the numerical results of the EE correlations of the halo major axes at $z=0$ 
to the linear model proportional to $\xi^{2}(r)$. The linear model with the 
best-fit amplitude is shown as dashed line in the bottom-right panel of 
Fig.\ref{fig:corz}. As can be seen, the linear model drops with $r$ too 
rapidly to fit the large-scale EE correlations.

We also fit the EE correlations of the halo minor axes measured at four 
redshifts and the EE correlations of the major axes measured from two 
different mass bins at $z=0$ to eq.~(\ref{eqn:cor}) and plot the results 
as solid lines in Figs.~\ref{fig:corz3} and \ref{fig:corm}.
\begin{table}
\centering
\caption{Mass bin, halo mean mass ($\bar{M}$) in unit of 
$10^{10}h^{-1}M_{\odot}$, and the best-fit values of $a$ for the EE 
correlations of the halo major axes.\label{tab:corm_fit}}
\begin{tabular}{@{}ccc@{}}
\hline
bin&$\bar{M}$ & $a\times 10^{2}$ \\
& [$10^{10}h^{-1}M_{\odot}$] & \\
\hline
low-mass  & $545.7$ & $0.86\pm 0.32$ \\
high-mass & $4915.6$ & $2.75\pm 0.39$ \\ 
\hline
\end{tabular}
\end{table}
The EE cross-correlation of the halo major axes between different mass 
bins at $z=0$ is similarly modeled as:
\begin{equation}
\label{eqn:ccor}
\eta_{C}(r) \approx  a_{c}\tilde{\xi}_{C}(r). 
\end{equation}
Here $a_{c}$ is a fitting parameter and $\tilde{\xi}_{C}(r)$ is defined as 
\begin{equation}
\label{eqn:cxi}
\tilde{\xi}_{C}(r) \equiv 
\frac{\int P(k)[\sin kr/kr]W(k;M_{1})W(k;M_{2})\,{\rm d}^{3}k}
{\int P(k)W(k;M_{1})W(k;M_{2})\,{\rm d}^{3}k}.
\end{equation}
where $M_{1}$ and $M_{2}$ represent the mean mass averaged over the low-mass 
and the high-mass bin, respectively. The best-fit value of $a_{c}$ is 
determined similarly by minimizing the $\chi^{2}$ function. 
The fitting result for the EE cross-correlation is plotted as solid line in 
Fig.~\ref{fig:corc}. As can be seen, the agreements between the numerical 
results and the fitting models are quite good for all cases.

\section{The Ellipticity-Direction Cross Correlations of Halos}

\subsection{Definition}

As mentioned in Section~1, another important correlation function for 
quantifying the cosmic web is the ED cross-correlation between the
halo ellipticities and the large-scale density field. If the halo 
ellipticities are induced by the anisotropic infall and merging along
the local filaments, then the orientations of the halo major axes must
be preferentially aligned with the directions to the neighboring
halos. This effect can be measured in terms of the ED cross-correlations
between the halo orientations and the location of halo neighbors.
We employ the following definition of the ED cross-correlations as 
\begin{equation}
\label{eqn:3rd}
\omega(r) \equiv \langle\vert\hat{\bf e}({\bf x})\cdot
\hat{\bf r}({\bf x})\vert^{2}\rangle - \frac{1}{3}, 
\end{equation}
where $\hat{\bf r}\equiv {\bf r}/r$ is a unit vector in the direction to 
a neighboring halo at separation distance of $r$. 

From the halo catalogs of the Millennium Run simulation at $z=0,0.5,1$ 
and $2$, we have measured the ED cross-correlations of the halo major 
($\omega_{\rm I}$), intermediate ($\omega_{\rm II}$) and minor
axes ($\omega_{\rm III}$) as a function of the comoving distance
$r$ between the halo centers. Basically, for each halo in the
Millennium data at a given redshift, we find the direction to its
neighbor halo and measure the separation distance $r$, and 
calculate the squares of the dot products of the normalized
eigenvector with the unit vector in the direction to the neighbor
halo. And then, we bin the radial distance $r$ and calculate the mean
values of $\vert\hat{\bf e}({\bf x})\cdot\hat{\bf r}({\bf x})\vert^{2}$ 
averaged over those halos whose distances to the neighbor halos belong 
to a given bin, subtracting $1/3$ from it. We perform this procedure at 
$z=2,1,0.5$ and $0$. 

\subsection{Evolution with redshift}

\begin{figure}
\begin{center}
\includegraphics[width=0.95\hsize]{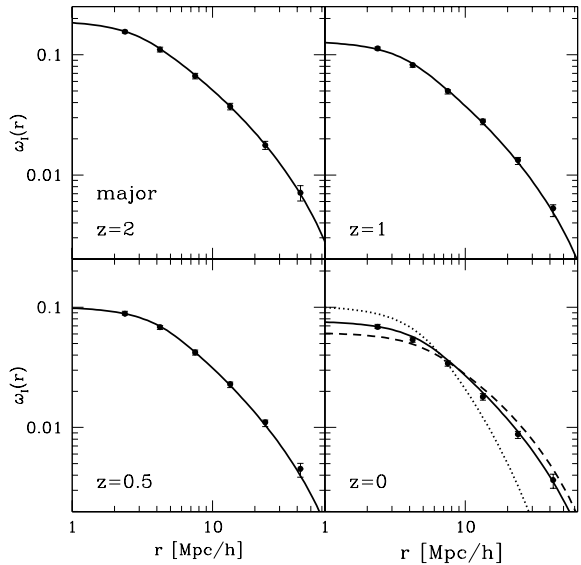}
\end{center}
\caption{The ED correlations of the halo major axes (solid dots) at
$z=0$, $0.5$, $1$ and $2$. The errors represent the standard deviation 
between realizations but do not include non-negligible correlations between 
radial bins. In each panel the solid line represents the best-fit model 
with two parameters (eq.~\ref{eqn:dir}). In the bottom-right panel 
the dotted and dashed line correspond to the best-fit results based on the 
model with one parameter proportional to $\xi$ and $\xi^{1/2}$, respectively. 
\label{fig:dirz}}
\end{figure}

Figure \ref{fig:dirz} plots $\omega_{\rm I}$ at redshifts $z=0$,
$0.5$, $1$ and $2$ in the top-right, top-left, bottom-right and
bottom-left panel, respectively. The errors represent the standard 
deviation between $8$ realizations. Since we are mainly interested 
in the cross-correlations between the halo principal axes and the 
large scale density field, we focus on separation scales greater than 
$1h^{-1}$Mpc.  As can be seen, the ED cross-correlations of the halo 
major axes also decrease as $z$ decreases at all distance scales. 
Note also that the ED cross-correlations are much stronger than the EE 
correlations shown in Fig.~\ref{fig:corz}.  The ED signal is statistically 
significant even at distances out to $50\,h^{-1}\,{\rm Mpc}$.

\begin{figure}
\begin{center}
\includegraphics[width=0.95\hsize]{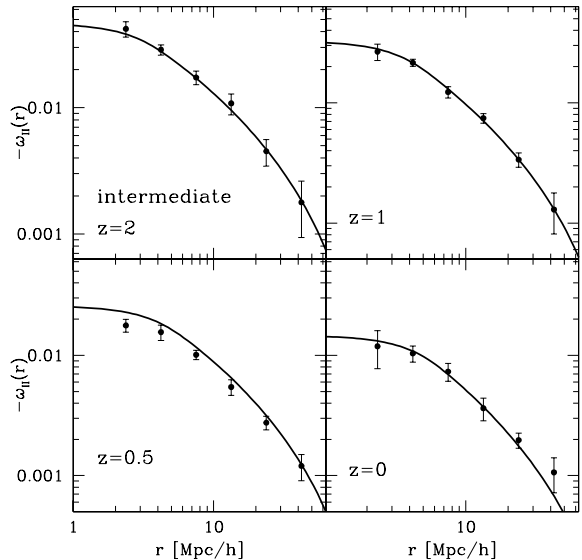}
\end{center}
\caption{Same as Fig.~\ref{fig:dirz} but for the case of the halo
intermediate axes.\label{fig:dirz2}}
\end{figure}

\begin{figure}
\begin{center}
\includegraphics[width=0.95\hsize]{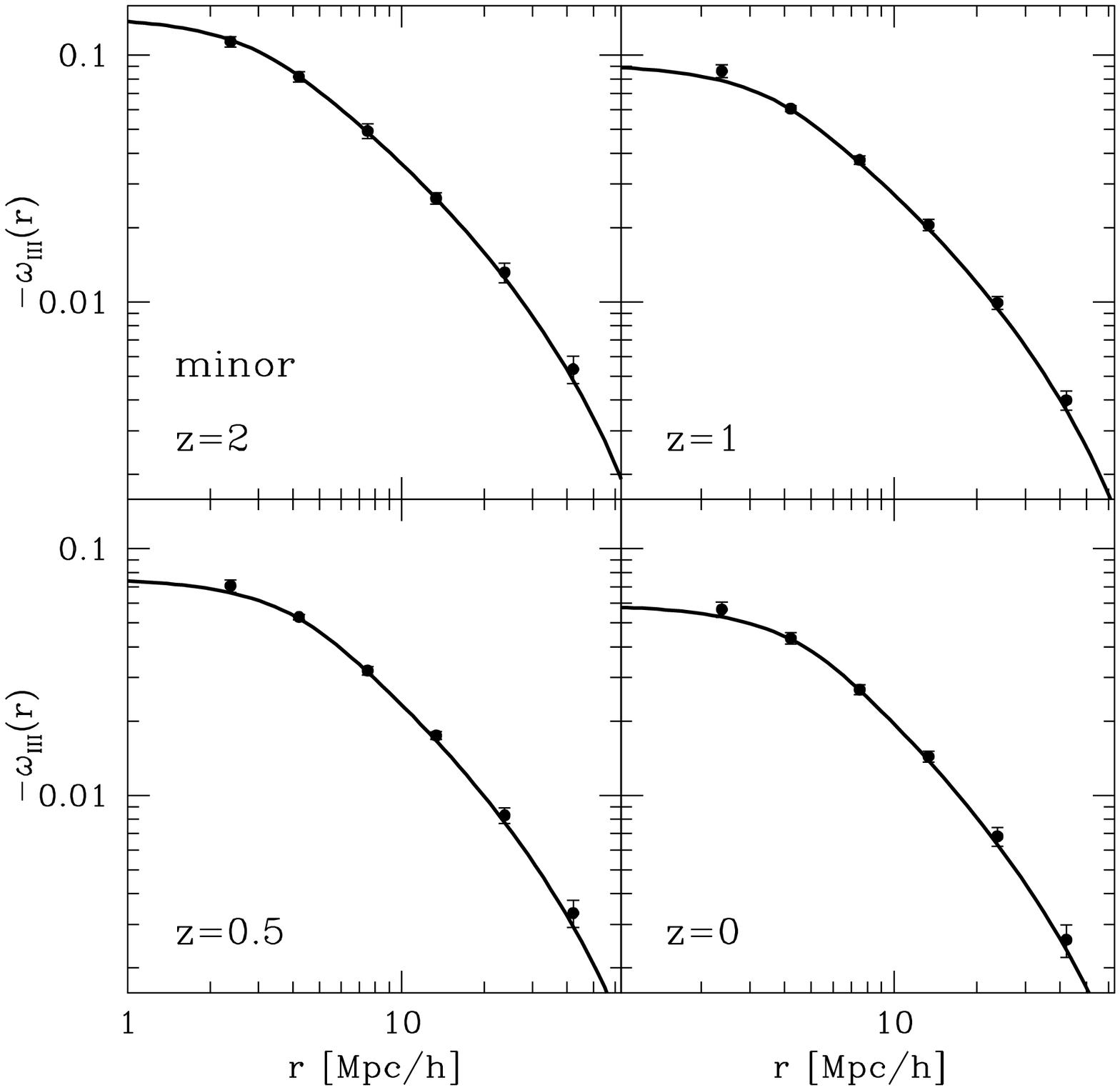}
\end{center}
\caption{Same as Fig.~\ref{fig:dirz} but for the case of the halo minor
axes.\label{fig:dirz3}}
\end{figure}

The ED cross-correlations of the intermediate and minor axes of halos,
$\omega_{\rm II}$ and $\omega_{\rm III}$, are plotted in Figs.~\ref{fig:dirz2} 
and \ref{fig:dirz3}, respectively. As expected, the intermediate and minor 
axes are {\it anti-correlated} with the directions to neighboring halos, and 
the degree of the anti-correlation is stronger for the minor axes. In fact, 
the ED anti-correlations of the halo minor axes are almost as strong as the 
ED correlations of the halo major axes. These results demonstrate clearly 
that the halo major axes preferentially point in the directions where the 
local density stays high, hence this gives a quantitative measure for the
filamentary distribution of the halos in the cosmic web. 
The numerical results on $\omega_{\rm I}$, $\omega_{\rm II}$ and 
$\omega_{\rm III}$ measured at z=0 are summarized in Table \ref{tab:ED_z0}. 

\begin{table}
\centering
\caption{Numerical results for the ED cross-correlations of the halo 
major ($\omega_{\rm I}$), intermediate ($\omega_{\rm II}$), and minor
($\omega_{\rm III}$) axes in logarithmic scale measured at $z=0$.
\label{tab:ED_z0}}
\begin{tabular}{@{}cccc@{}}
\hline
$\log [r/(h^{-1}{\rm Mpc})]$ & $\omega_{\rm I}(r)\times 10^{2}$ & 
$-\omega_{\rm II}(r)\times 10^{2}$ & $-\omega_{\rm III}(r)\times 10^{2}$\\
\hline
$0.375$ & $6.85\pm 0.35$ & $1.19\pm 0.42$ & $5.67\pm 0.40$\\
$0.625$ & $5.37\pm 0.26$ & $1.04\pm 0.16$ & $4.33\pm 0.22$\\
$0.875$ & $3.42\pm 0.17$ & $0.73\pm 0.12$ & $2.68\pm 0.12$\\
$1.125$ & $1.80\pm 0.11$ & $0.36\pm 0.08$ & $1.43\pm 0.07$\\
$1.375$ & $0.88\pm 0.05$ & $0.20\pm 0.03$ & $0.68\pm 0.06$\\
$1.625$ & $0.37\pm 0.04$ & $0.11\pm 0.03$ & $0.26\pm 0.04$\\
\hline
\end{tabular}
\end{table}

\subsection{Variation with mass}

\begin{figure}
\begin{center}
\includegraphics[width=0.95\hsize]{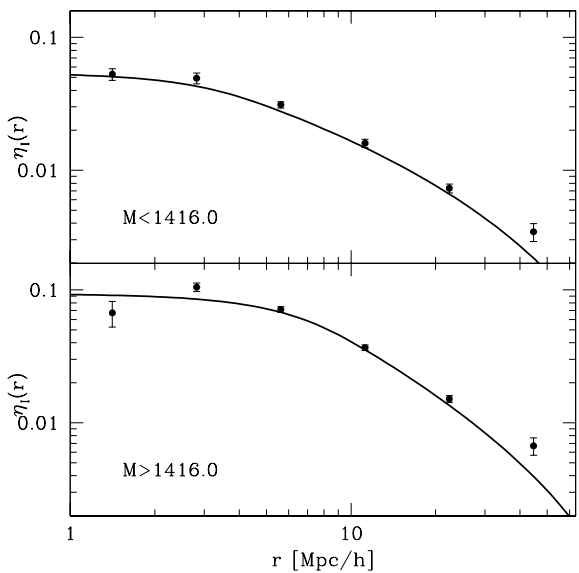}
\end{center}
\caption{The ED cross-correlations of the halo major axes measured from the 
low-mass ($M<M_{c}$) and the high-mass ($M>M_{c}$) bins at $z=0$ in the top 
and the bottom panel, respectively. The mass threshold $M_{c}=1416$ is 
in unit of $10^{10}h^{-1}M_{\odot}$.
\label{fig:dirm}.}
\end{figure}

To explore how the ED correlation changes with halo mass, we
measure the ED correlation of the halo major axes at two 
different mass bins with the mass threshold $M_{c}=1.42\times 
10^{13}h^{-1}M_{\odot}$ at $z=0$. . When finding the neighbors, 
we consider all halos, no matter what mass the neighbor halos have. 
Figure \ref{fig:dirm} plots the ED correlations of the halo major axes
at two different mass bins at $z=0$. As can be seen, the ED 
correlations increase as the halo mass increases, just like the 
EE correlations, which suggests that the anisotropic merging 
contributes significantly to the ED correlations.

\subsection{Fitting formula}

Given the observed fact that the EE correlations scale linearly with
the linear density two-point correlation function (Section~3.4), we
have also tried in vain to model the ED correlations as a linear
scaling of the density correlation function. But, it has turned out
that this simple model fails in providing good fits to the large-scale 
ED cross correlations (see Fig.~\ref{fig:dirz}).  To improve the fitting 
result, we need a model which decreases with $r$ more slowly than $\xi(r)$. 
We suggest the following empirical formula:
\begin{equation}
\label{eqn:dir}
\omega(r)\approx b_{1}\tilde{\xi}_{A}(r) + b_{2}\tilde{\xi}^{1/2}_{A}(r)
\end{equation}
where the two parameters $b_{1}$ and $b_{2}$ lie in the range of
$[-1/3,\,2/3]$. The second term proportional to $\xi^{1/2}$ is included to 
fit the large-scale ED correlations. For the halo major-axis, both of 
$b_{1}$ and $b_{2}$ will have positive values, while for the halo minor and 
intermediate axes, they will be negative. We also expect that $b_{1}$ and 
$b_{2}$ will have larger values than $a$, since the ED cross-correlation is 
a more direct measure of the filamentary distribution of dark matter halos.

We fit the numerical results on the ED correlations obtained in 
Sections~4.2-4.3 to eq.~(\ref{eqn:dir}) and determine the best-fit 
values of $b_{1}$ and $b_{2}$ by minimizing the $\chi^{2}$ function:
\begin{equation}
\chi^{2} = [\omega_{i}-\omega(r_{i};b_{1},b_{2})]C^{-1}_{ij}
[\omega_{j}-\omega(r_{j};b_{1},b_{2})], 
\end{equation}
where $\omega_{i}$ is the numerical data point at the $i$-th distance bin,
$r_i$, and $\omega(r_{i};b_{1},b_{2})$ is the fitting model at $r_{i}$.
The covariance matrix, $(C_{ij})$, is calculated as 
\begin{equation}
\label{eqn:cov_ome}
C_{ij} = \langle(\omega_{i}-\omega_{0i})(\omega_{j}-\omega_{0j})\rangle, 
\end{equation} 
where $\omega_{0i}$ is the mean $\omega_{i}$ obtained from the whole 
simulation box.  To calculate errors in the measurement of $b_{1}$ and 
$b_{2}$, we first construct a $2\times 2$ curvature matrix defined as 
\begin{equation}
{\cal F}_{ij} = 
\left(\frac{\partial^{2}\chi^{2}}{\partial b_{i}\partial b_{j}}\right), 
\end{equation}
with $i,j=\{1,2\}$. The errors are calculated as the diagonal components of
the inverse curvature matrix \citep{dod03}:
\begin{equation}
\sigma^{2}_{b_1} = \left({\cal F}^{-1}\right)_{11}, \qquad 
\sigma^{2}_{b_2} = \left({\cal F}^{-1}\right)_{22}.
\end{equation}

Table \ref{tab:dirz} lists the best-fit values of $b_1$ and $b_2$
for the ED correlations of the halo major axes measured at
$z=0,0.5,1$ and $2$. The best-fit value of the parameter $b_2$ is
larger than that of $b_1$ at every redshift, indicating that the second
term in eq.~(\ref{eqn:dir}) dominates. The values of the two
parameters decrease monotonically as $z$ decreases, just like the EE
correlations. Our results provide a quantitative description of the 
evolution of the ED correlations and its scaling with distance. 

The fitting results on the ED correlations of $\omega_{\rm I}$, 
$\omega_{\rm II}$ and $\omega_{\rm III}$ at $z=0$ with the best-fit values 
of $b_1$ and $b_2$ are plotted as solid lines in Figs.~\ref{fig:dirz}, 
\ref{fig:dirz2}, and \ref{fig:dirz3}, respectively. As can be seen, the 
agreements between the fitting models and the numerical results are quite 
good at all redshifts. For comparison, we also fit the numerical results of 
the ED correlations of the halo major axes at $z=0$ to two different models 
proportional to $\xi$ and $\xi^{1/2}$, which are plotted in the bottom right 
panel as dotted and dashed line, respectively. As can be seen, the model 
proportional to $\xi$ decreases with $r$ too rapidly to fit the numerical 
results. Meanwhile the $\xi^{1/2}$ decreases slowly with $r$ but it alone 
still does not agree with the numerical results as well as 
eq.~(\ref{eqn:dir}). 

\begin{table}
\centering
\caption{The best-fit values of the two correlation parameters for the
ED correlations of the halo major axes at four different redshifts.
\label{tab:dirz}}
\begin{tabular}{@{}cccc@{}}
\hline
z & $b_{1}\times 10^{2}$ & $b_{2}\times 10^{2}$ & \\ \hline
 0   & $3.06\pm 0.20$ & $4.61\pm 0.11$ & \\
 0.5 & $4.54\pm 0.21$ & $5.55\pm 0.12$ & \\
 1   & $5.73\pm 0.23$ & $7.25\pm 0.10$ & \\
 2   & $7.57\pm 0.32$  & $11.64\pm 0.16$ & \\ 
\hline
\end{tabular}
\end{table}

Figure \ref{fig:dirm} plots the fitting results on the ED correlations of 
the major axes of the low-mass and high-mass halos at $z=0$ in the top and 
bottom panel, respectively. Table \ref{tab:dirm} lists the best-fit values 
of $b_{1}$ and $b_{2}$ for the two cases. The values of $b_{1}$ and $b_{2}$ 
are higher for the high-mass halos than for the low-mass halos. 
Note that although the fitting models work fairly well, they seem to deviate 
from the numerical results by more than $3\sigma_{\omega}$ around 
$30\,h^{-1}\,{\rm Mpc}$. We think that it reflects the failure of the 
assumption that the ED correlations can be described in terms of 
the linear density correlation function. 
\begin{table}
 \centering
  \caption{The best-fit parameters of the ED cross-correlations 
measured from two different mass bins at $z=0$.\label{tab:dirm}}
  \begin{tabular}{@{}cccc@{}}
  \hline
mass bin & $\bar{M}$ & $b_{1}\times 10^{2}$ & $b_{2}\times 10^{2}$ \\
& [$10^{10}h^{-1}M_{\odot}$] &  &  \\ \hline
low-mass & $545.7$  & $1.39\pm 0.36$ &  $4.02\pm 0.14$ \\
high-mass & $4914.35$ & $4.94\pm 0.29$  & $4.39\pm 0.16$ \\
\hline
\end{tabular}
\end{table}

\section{Summary and Discussion}

In this work, by analyzing the halo data and the semi-analytic galaxy
catalog from the Millennium simulations at $z=0$, $1$, $0.5$ and $2$,
we have measured the ellipticity-ellipticity (EE) correlations. 
The correlations are close to $0.01$ at a distance of $1\,h^{-1}\,{\rm
Mpc}$ and remain significant at distances out to $10\,h^{-1}\,{\rm
Mpc}$. The EE correlations are found to be strongest for the case of 
the halo major axes, and weakest for the case of the intermediate axes. 

We have found that the EE correlations of all three axes decrease as
$z$ decreases. This might be due to the growth of secondary filaments 
at low redshifts and the beginning `freeze-out' of structure growth in 
$\Lambda$CDM, which plays a role in randomizing the halo ellipticities.
It has been also  found that the EE correlation function exhibits a
strong dependence on halo mass. It increases as the mass scale
increases, which might be due to the dominant filamentary merging of
halos on large scales. We have also calculated EE cross-correlations 
between halos belonging to different mass bins. Our results have shown 
that the EE cross-correlations between neighboring mass bins exist at a 
statistically significant level as well.

We have modeled the EE correlation function as a linear scaling of the
linear density two-point correlation function $\xi(r)$, which is characterized
by one fitting parameter, $a$. The value of $a$ represents the
amplitude of the EE correlations, quantifying its scaling with mass
and redshift. The fitting model with the best-fit value of $a$ has
been shown to agree with the numerical results quite well at all
redshifts and all mass bins.

We have also measured the cross-correlations between the halo
principal axes and the directions to neighboring halos (ED) by using
the same numerical data, and found that the ED cross-correlations are
much stronger than the EE correlations, at all distances. 
Remarkably, they are detected even at distances out to $50\,h^{-1}{\rm
Mpc}$ at a statistically significant level. Just like the EE
correlations, the ED cross-correlations are found to decrease as $z$ 
decreases and increase as the halo mass $M$ increases, suggesting a
dominant role of anisotropic merging and infall of matter in
establishing these correlations. The intermediate and the minor axes of the
halos have turned out to be anti-correlated with the directions to the 
neighboring halos, which is consistent with alignments of the halos 
shapes with the orientations of the local filament.

The ED cross-correlations are, however, found to be poorly fitted by a
linear scaling of $\xi(r)$. The ED cross-correlations
decrease with distance much less rapidly than $\xi(r)$. To account for
the slow decrease of ED cross-correlations with distance, we include
an additional term proportional to $\xi^{1/2}(r)$ in the fitting model
which is then characterized by two fitting parameters $b_!$ and $b_2$. 
Thetwo parameters represent the amplitudes of the two terms of the ED
cross-correlations proportional to $\xi(r)$ and $\xi^{1/2}(r)$,
respectively. This fitting formula has been shown to agree with the
numerical results quite well at all redshifts and at all mass bins.

Nonetheless, it is worth mentioning here that our fitting formula for
the ED cross-correlations is not a physical model. It is purely
empirical, obtained through comparison with the numerical results. It
has yet to be understood why the ED cross-correlations scale as
described by our fitting formula. At any rate, we believe that our
fitting formula may be useful in the future when the ED
cross-correlations can be modeled by a fundamental theory.

The EE correlations and the ED cross-correlations that we have
measured here provide a useful tool to statistically characterize the
anisotropy and the relevant scales of the cosmic web. It will be
interesting to compare the results we obtained here for the
$\Lambda$CDM cosmology with observational data from large galaxy
redshift surveys. The comparison of our numerical results with
observational data, however, will require a modelling of the redshift
space distortions as well as of two dimensional projection effects,
given that in real observations what can be usually measured is the
two dimensional projected major axes of the galaxies in redshift
space. In future work, we plan to model these two effects on the EE
and ED correlations and compare the numerical results with real
observational data.

Another important application of our results lies in studies of weak
gravitational lensing.  The issue of a potential cross-correlation
between galaxy ellipticities and the weak gravitational lensing shear
(GI cross-correlations) was first raised by \citet{hir-sel04}. They
claimed that if such GI cross-correlations exist, then they would affect
the weak lensing signal as another systematic contaminant whose effect
is hard to control. The GI cross-correlations are expected to occur
primarily due to the ED cross-correlations: If the intrinsic
ellipticities of the galaxies are cross-correlated with the surrounding
large-scale density field, then it will in turn lead to a cross-correlation 
between the gravitational lensing shear and the galaxy ellipticities. 
Recent observations indeed have reported detections of the GI
correlation signals in low-redshift galaxy surveys 
\citep{man-etal06,hir-etal07}. To assess a possible systematic
contamination of weak lensing due to the GI cross-correlations, 
it will be important first to examine the relation between the
observed GI cross-correlations and the ED cross-correlations of the 
cosmic web. This work will require incorporating a model for how the 
galaxy shapes are aligned relative to the dark matter
\citep{hey-etal06}. Our future work is in this direction. 

\section*{Acknowledgments}

The Millennium Simulation analyzed in this paper was carried out by the
Virgo Supercomputing Consortium at the Computing Center of the
Max-Planck Society in Garching, Germany. The simulation databases 
and the web application providing online access to them were
constructed as part of the activities of the German Astrophysical
Virtual Observatory.G.L. works for the German Astrophysical Virtual 
Observatory (GAVO),which is supported by a grant from the German
Federal Ministry of Education and Research (BMBF) under contract 05 AC6VHA.

We thank an anonymous referee for his/her constructive report which 
helped us improve significantly the original manuscript. We also thank 
S.D.M.~White for stimulating discussion and useful suggestions.
J.L. is very grateful to the warm hospitality of the Max Planck Institute for 
Astrophysics (MPA) in Garching where this work was initiated and performed. 
J.L. acknowledges financial support from Korea Science and Engineering 
Foundation (KOSEF) grant funded by the Korean Government (MOST,
NO. R01-2007-000-10246-0).

\bsp

\label{lastpage}

\end{document}